\documentclass[pdflatex,sn-nature]{sn-jnl}

\usepackage{graphicx}%
\usepackage{multirow}%
\usepackage{amsmath,amssymb,amsfonts}%
\usepackage{amsthm}%
\usepackage{mathrsfs}%
\usepackage{comment}
\usepackage[title]{appendix}%
\usepackage[dvipsnames]{xcolor}%
\usepackage{textcomp}%
\usepackage{manyfoot}%
\usepackage{booktabs}%
\usepackage{algorithm}%
\usepackage{algorithmicx}%
\usepackage{algpseudocode}%
\usepackage{listings}%
\theoremstyle{thmstyleone}%
\theoremstyle{thmstyletwo}%

\theoremstyle{thmstylethree}%

\begin{document}

\title[Direct imaging of sub-parsec binary SMBHs in the central galaxy of the NGC 5044 group]{Direct imaging of sub-parsec binary SMBHs in the central galaxy of the NGC 5044 group}


\author*[1,2]{\fnm{Francesco} \sur{Ubertosi}}\email{francesco.ubertosi2@unibo.it}

\author[3]{\fnm{Gerrit} \sur{Schellenberger}}%
\author[3]{\fnm{Ewan} \sur{O'Sullivan}}%

\author[3]{\fnm{Laurence P.} \sur{David}}

\author[3]{\fnm{William} \sur{Forman}}

\author[4]{\fnm{Simona} \sur{Giacintucci}}

\author[3]{\fnm{Christine} \sur{Jones}}

\author[3]{\fnm{Kamlesh} \sur{Rajpurohit}}

\author[2]{\fnm{Tiziana} \sur{Venturi}}

\author[3]{\fnm{Jan} \sur{Vrtilek}}

\affil*[1]{\orgdiv{Dipartimento di Fisica e Astronomia}, \orgname{Università di Bologna}, \orgaddress{\street{via Gobetti 93/2}, \city{Bologna}, \postcode{I-40129}, \country{Italy}}}

\affil[2]{\orgdiv{Istituto di Radioastronomia (IRA)}, \orgname{Istituto Nazionale di Astrofisica}, \orgaddress{\street{via Gobetti 101}, \city{Bologna}, \postcode{I-40129}, \country{Italy}}}

\affil[3]{\orgdiv{Center for Astrophysics $\rvert$ Harvard \& Smithsonian}, \orgaddress{\street{60 Garden Street}, \city{Cambridge}, \postcode{MA 02138}, \country{USA}}}

\affil[4]{\orgdiv{Naval Research Laboratory}, \orgaddress{\street{4555 Overlook Avenue SW, Code 7213}, \city{Washington}, \postcode{DC 20375}, \country{USA}}}


\abstract{
Binary active galactic nuclei (AGN) consist of two supermassive black holes (SMBHs) hosted in a single galaxy that are simultaneously accreting gas. These systems are key to understanding hierarchical galaxy evolution and represent prime targets for low-frequency gravitational wave (GW) experiments. However, binary AGN remain extremely rare: only a handful are confirmed at separations of a few hundred parsecs, and just one pc-scale binary - in an elliptical galaxy at the center of a galaxy cluster - has been securely and visually identified. Direct imaging of sub-parsec binary AGN, the final stage before the GW regime, have so far been missing. Here we report on Very Long Baseline Array (VLBA) observations of NGC 5044, the central galaxy of the X-ray brightest galaxy group. This galaxy is known to exhibit pc-scale radio jets nearly orthogonal to two older AGN outbursts. The 4.9, 8.4, 15.2, and 23.6 GHz VLBA images reveal two compact, flat-spectrum, and variable radio cores separated by just 0.38 parsec at the heart of the galaxy, forming the tightest directly imaged binary AGN known by an order of magnitude. This discovery suggests that the older outbursts may have been powered by jets from the secondary core, without invoking any reorientation of the primary jet axis. More broadly, comparison with other systems points to massive, cluster-central elliptical galaxies being promising targets to search for binary SMBHs.
}


\maketitle

\section{Introduction}\label{sec:intro}
In hierarchical models of galaxy formation, galaxies grow through successive mergers (e.g., \cite{Baugh2006}). Since supermassive black holes (SMBHs) are ubiquitous in galaxies \cite{Kormendy1995}, mergers naturally produce SMBH pairs that sink to the galactic center via dynamical friction, leaving a fraction of galaxies hosting bound binary SMBHs \cite{Volonteri2003}. The occurrence rate of these binaries constrains galaxy and SMBH merger timescales, key ingredients of cosmological simulations of galaxy evolution (e.g., \cite{Volonteri2003,BurkeSpolaor2011,RodriguezGomez2015}). Additionally, as SMBH binaries reach sub-pc separations, they become emitters of nHz gravitational waves (GW). Therefore, the demographics (i.e., occurrence rates) of milli-pc and sub-pc binaries, as well as of their progenitors (pc-scale binaries) is crucial for modeling the nHz stochastic GW background that pulsar timing array (PTA) experiments have revealed (e.g., \cite{DeRosa2019,Comerford2025}). 
Despite extensive efforts, observational constraints on pc-scale, bound SMBH binaries remain extremely limited \cite{DeRosa2019,Severgnini2022,DAmato2026}. Optical spectroscopy can be a useful tool to probe binaries at sub-pc/pc separations, since double-peaked broad emission lines can trace gas bound to each SMBH (e.g., \cite{Boroson2009,Eracleous2012}). However, similar line profiles can also arise from other processes, such as disk emission from a single AGN or gas outflows, making spectroscopic candidates uncertain without independent confirmation (e.g., \cite{Runnoe2017,Doan2020,Breiding2021,DOrazio2023}). The only direct method to image and confirm a bound binary requires Very Long Baseline Interferometry (VLBI) imaging, capable of resolving two distinct compact cores at the sub-pc to pc scale. To date, only one system, the galaxy 4C+37.11, has been confirmed with VLBI observations to host a binary SMBHs system with separation $d_{sep}=7.3$~pc \cite{Rodriguez2006,Bansal2017}. 
Although a few other candidates of varying binary separations have been proposed from VLBI imaging (e.g., \cite{Gitti2013,Deane2014,Kharb2017,An2018}), most were later ruled out by deeper, multi-frequency follow-up. The closest-separation claim was NGC~7674, with a proposed $d_{sep}\sim0.35$~pc binary based on a single-frequency, $\sim5\sigma$ detection of a secondary core \cite{Kharb2017}; subsequent deeper and multi-frequency VLBI imaging found no evidence for a second core, ruling out the binary \cite{Breiding2022}. Similarly, a $d_{sep}\sim77$~pc candidate in RBS~797, also based on a single-frequency $\sim5\sigma$ detection \cite{Gitti2013}, was excluded by higher-sensitivity, multi-frequency VLBI data \cite{Ubertosi2024_RBS797}. These examples highlight the need for deep, multi-frequency, and possibly multi-epoch data to robustly investigate the presence of binary AGN in the radio, as was the case for 4C+37.11 \cite{Rodriguez2006,Bansal2017}.
Equally important is identifying where to search for these systems: models predict that bound SMBH binaries are most likely to be found in massive elliptical galaxies, particularly the brightest central galaxies (BCGs) of clusters and groups, which have grown to stellar masses of $M_{\ast}\sim10^{12}$~M$_{\odot}$ through multiple mergers, and host the most massive SMBHs (e.g., \cite{Volonteri2003,Hoffman2007}). 
These galaxies are also the most likely to host radio-loud active galactic nuclei (AGN) \cite{Best2005,Sabater2019}, making them the optimal targets for a radio VLBI search. Notably, 4C+37.11 itself is the BCG of a local, relaxed galaxy cluster \cite{Romani2014,andradesantos2016}, the system 1RXS~J040547.3+380308 \cite{Voges1999}. \\

\par In this work, we report the discovery of a sub-parsec SMBH binary in the elliptical galaxy NGC~5044 ($z = 0.009$), the central massive galaxy of the X-ray brightest galaxy group in the sky \cite{Buote2003}, 
and probably the best-studied example of SMBH feedback in galaxy groups (e.g., \cite{Gastaldello2009,David2014,David2017,schellenberger2020,Schellenberger2021,Ubertosi2024,Rajpurohit2025,Ubertosi2026,Temi2026,Tamhane2026}).  
We assume a $\Lambda$CDM cosmology with $H_{0} = 70$~km~s$^{-1}$/Mpc, $\Omega_{m} = 0.3$, and $\Omega_{\Lambda} = 0.7$. The spectral index $\alpha$ is defined as $S_{\nu}\propto\nu^{\alpha}$, where $S_{\nu}$ is the flux density at frequency $\nu$. With a systemic velocity of 2757~km~s$^{-1}$ for NGC~5044 \citep{schellenberger2020}, and a distance of 31.2 Mpc \citep{Tonry2001}, 1~mas corresponds to  0.15~pc in the rest frame of NGC~5044.


\section{Results}\label{sec:results}
\par
\begin{figure}[htp!]
\centering
\includegraphics[width=\textwidth]{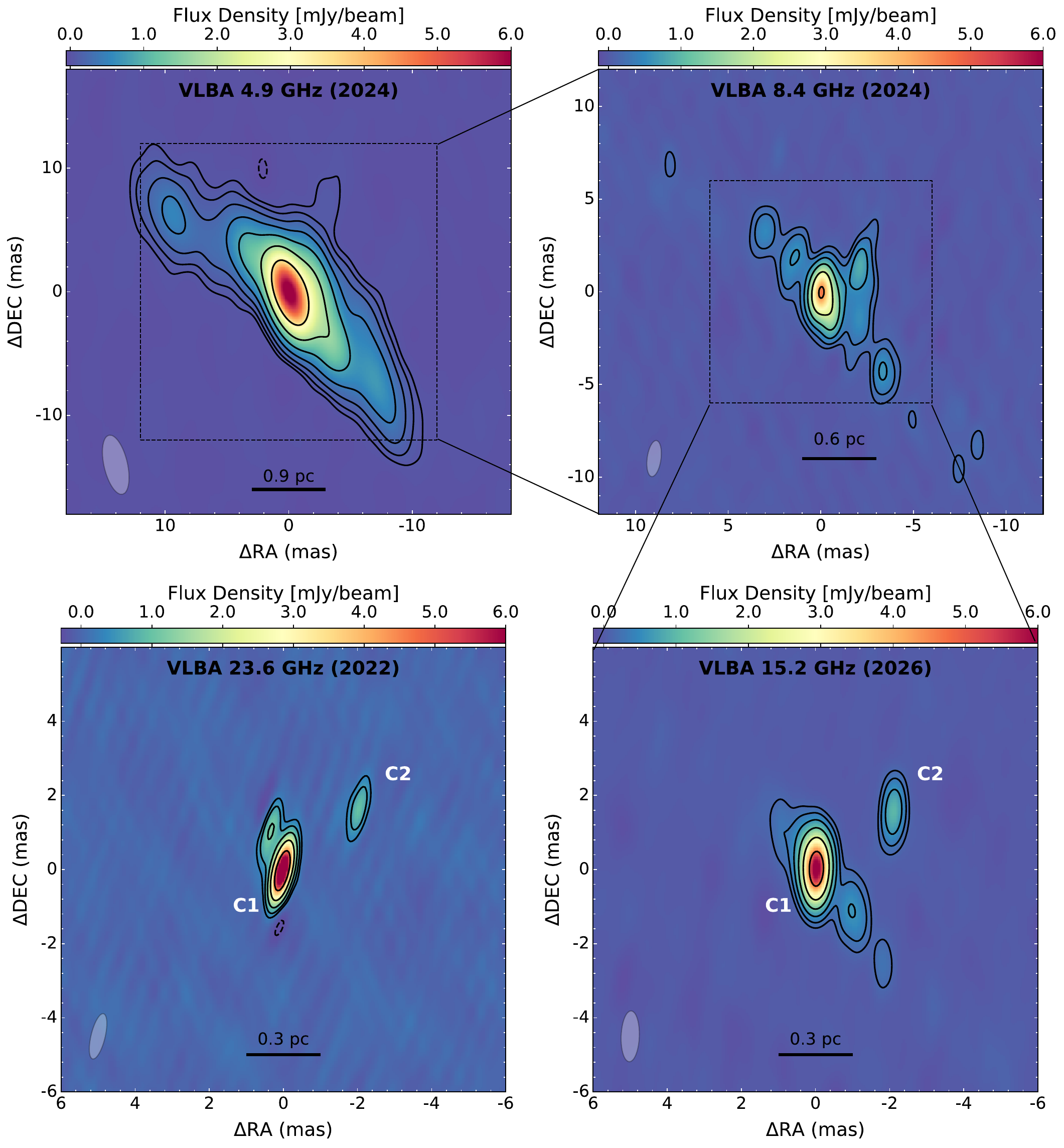}
\caption{Multi-wavelength VLBA images of the core of NGC~5044, on a linear color-scale. Clockwise from top left: 4.9~GHz, 8.4~GHz, 15.2~GHz, and 23.6~GHz. The bright central region, unresolved at 4.9~GHz, splits into two compact components at higher frequencies: the primary core (C1), from which the northeast-southwest jets originate, and a secondary core (C2) at $\sim$0.4~pc separation, labeled in the 15.2 and 23.6~GHz panels. Contours are drawn at $5\sigma_{rms}$, increasing by factors of two (imaging parameters are given in {\bf Methods}, Table~\ref{tab:imgparams}).
}\label{fig:imagesVLBA}
\end{figure}

 To investigate the radio core at the highest possible angular resolution, here we present multi-frequency VLBA images at 4.9~GHz, 8.4~GHz, 15.2~GHz, and 23.6~GHz, reaching an angular resolution of around 1~mas (0.15~pc; see \textbf{Methods}), a factor of $3\times$ sharper than previous studies of NGC~5044 (\cite{Schellenberger2021,Ubertosi2024,Ubertosi2026}). The multi-frequency images shown in 
Figure \ref{fig:imagesVLBA} clearly recover a northeast-southwest jet structure, organized in several knots of synchrotron emission, already known from previous studies (\cite{Schellenberger2021,Ubertosi2024,Ubertosi2026}). The bright central region, unresolved at 4.9 GHz, is resolved into two components in the 8.4 GHz, 15.2 GHz, and 23.6 GHz images: a brighter component (C1) that we identify as the primary core from which the northeast-southwest jets originate, and a second fainter component (C2) at a distance of $\sim$2.5~mas (0.4~pc) northwest of the primary core. This secondary component is detected at a signal-to-noise ratio (SNR) ranging from 10 to $\geq30$ (Figure~\ref{fig:imagesVLBA}; the SNR varies with frequency and rms noise, see also \textbf{Methods} and Table~\ref{tab:difmap}). 

\par
\begin{figure}[htp!]
    \centering
    \includegraphics[width=0.45\linewidth]{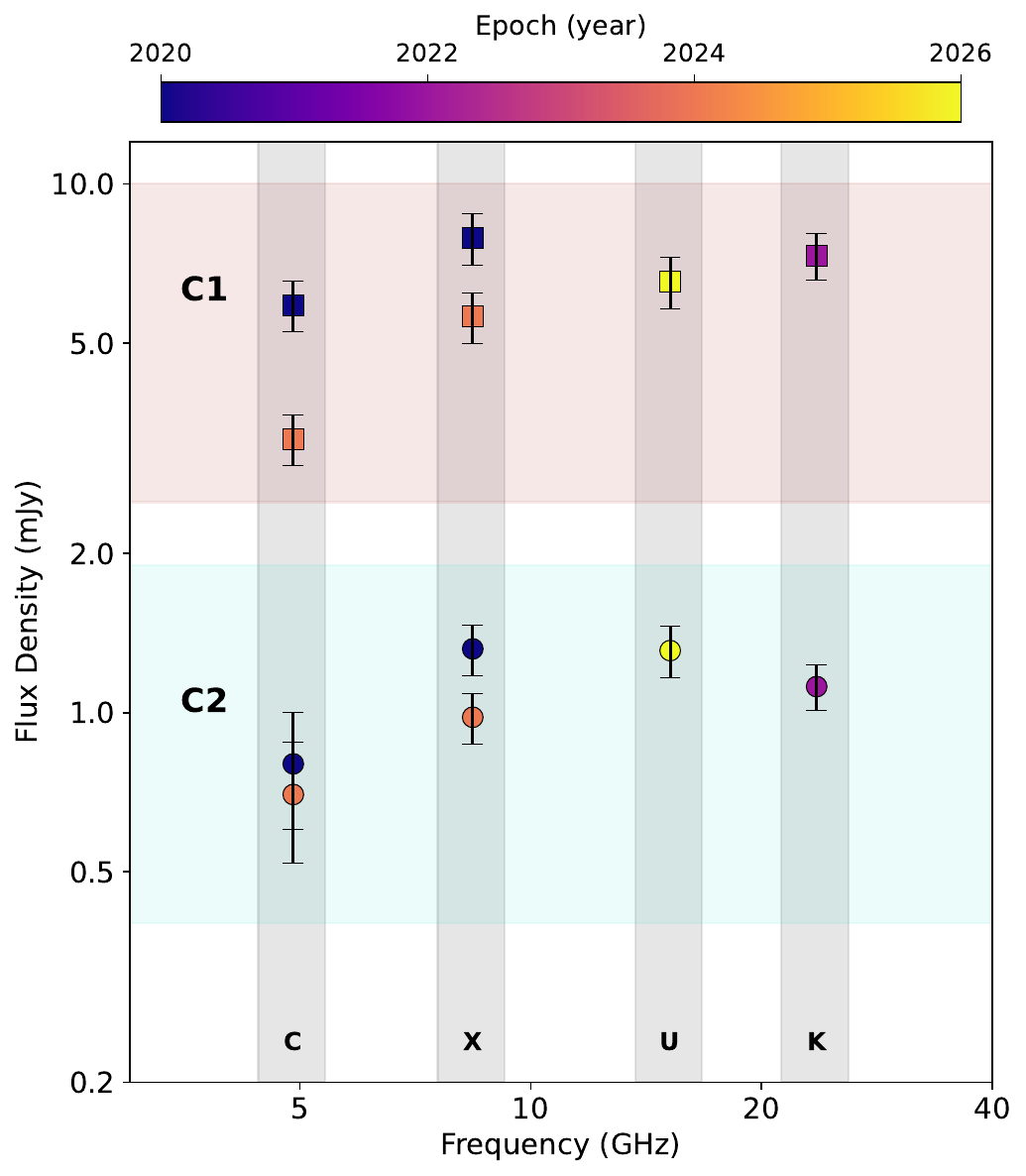}
    \includegraphics[width=0.45\linewidth]{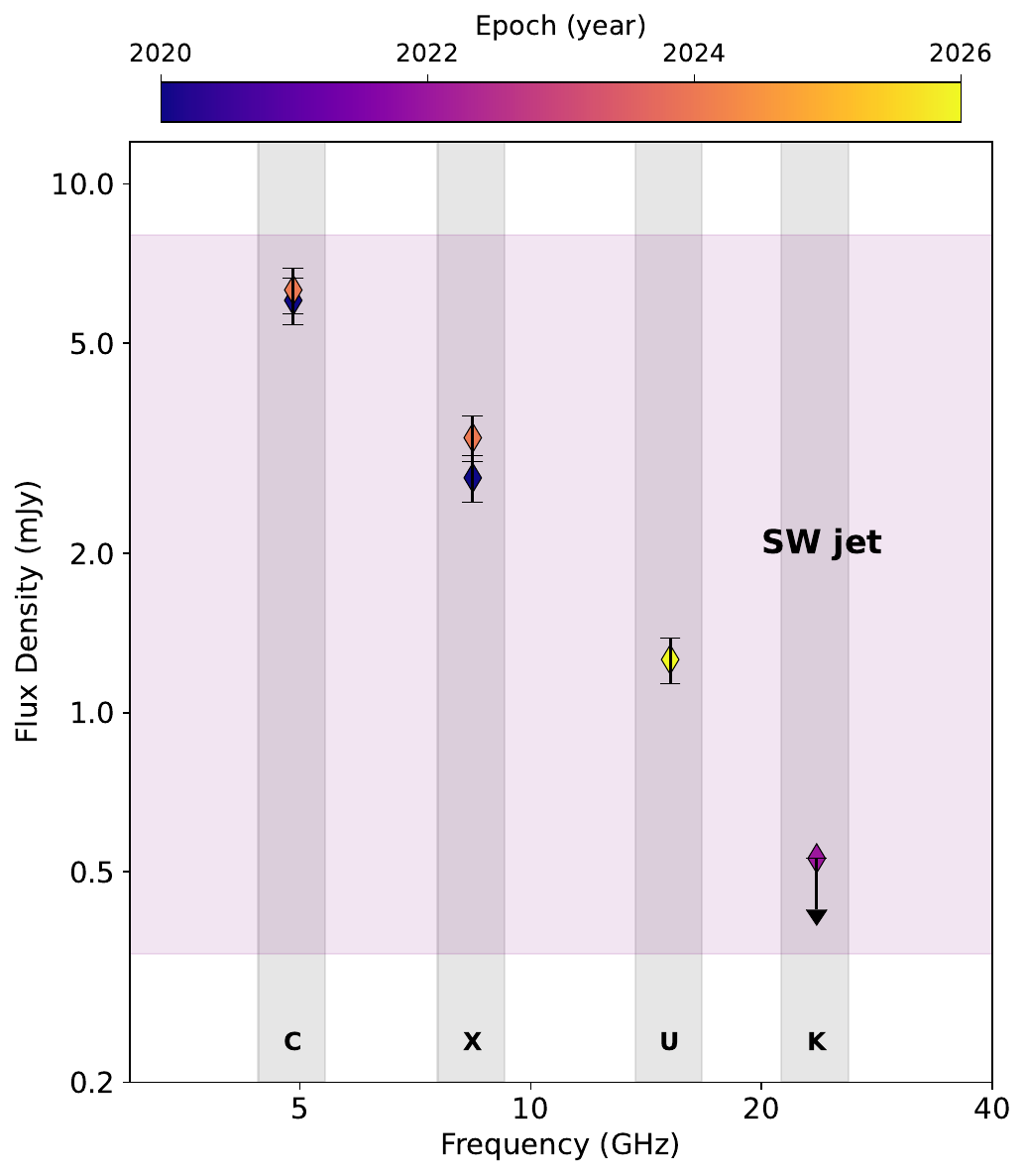}
    \caption{Radio spectra of the binary SMBHs C1 and C2 (left) and of the southwest jet knot (right), measured from the multi-frequency, multi-epoch dataset summarized in Table~\ref{tab:difmap}, color-coded by epoch of the observation. C1 and C2 show flat-to-inverted spectra typical of AGN cores, while the jet knot shows the negative spectral index typical of optically thin synchrotron emission (see {\bf Methods}). Note that the different colors at the two lowest frequencies refer to different epochs, while the shaded vertical gray regions show the frequency of the observing bands. The down-pointing arrow in the right panel represents an upper limit on the SW jet flux density at 23.6 GHz.  
    }
    \label{fig:spectracomponents}
\end{figure}

 In Figure~\ref{fig:spectracomponents} we show the integrated spectra of C1 and C2, obtained by measuring the flux density of these features from the datasets at different frequencies and epochs with uniform sensitivity and uv-coverage (see {\bf Methods} and Table~\ref{tab:difmap}). Across the different epochs, the flux densities of C1 and C2 have varied. Specifically, the flux density of C1 has decreased by $\sim$40\% at 4.9 GHz and by $\sim$30\% at 8.4 GHz from March 2020 to September 2024 (4.5~y), while the flux density of C2 has decreased by 20\% at 8.4 GHz over the same time interval. Both components show nearly flat or inverted spectra between 4.9 GHz and 23.6 GHz. Conversely, within the jet-knot structure, the feature we label southwest (SW) jet -- the knot of jet emission southwest of C1 and closest to the core -- is characterized by a negative, steep spectral index and no significant variability at low frequencies between 2020 and 2024; we estimate $\alpha_{4.9}^{8.4} = -1.2\pm0.3$, consistent with an optically thin AGN jet (e.g., \cite{Hovatta2014}). Based on our modeling of the source structure in the visibility domain (see \textbf{Methods}), C1 and C2 appear compact (unresolved) at all frequencies, with deconvolved sizes of $\leq0.4$ mas ($\leq0.06$ pc; see Table~\ref{tab:difmap}) -- smaller than the angular resolution. By contrast, the SW jet is best described as a spatially extended component (as visible in high angular resolution maps), with a deconvolved size of 0.7 - 2 mas (0.1 - 0.3 pc).

\par The properties of C1 and C2 -- specifically their compact radio emission, flat or inverted spectral indices, and flux variability over time -- are consistent with those of AGN cores. We rule out the possibility that C2 is a jet knot launched by C1, which is clearly producing jets in the northeast–southwest direction (Figure~\ref{fig:imagesVLBA}). After correcting for a (mild) core-shift (e.g., \cite{Lobanov1998}) effect (see {\bf Methods}), we measure a projected angular separation of $r_{proj} = 2.52\pm0.10$~mas, corresponding to $d_{sep} = 0.38\pm0.02$~pc. Considering these physical properties together with their projected physical separation, we conclude that NGC~5044 hosts a sub-pc SMBH binary directly imaged using the radio emission from the two cores. We note that, prior to this work, the closest-separation VLBI claim after 4C+37.11 ($d_{sep} = 7.3$~pc) was NGC~7674, with a proposed $d_{sep}\sim0.35$~pc binary based on a single-frequency detection of a secondary core \cite{Kharb2017}; this was later ruled out by deeper, multi-frequency VLBI imaging, which found no evidence for a second core \cite{Breiding2022}.
Therefore, the projected distance of C1 and C2 stands as the lowest ever measured for confirmed and candidate VLBI-imaged SMBH binaries (e.g., \cite{Rodriguez2006,Deane2014,Bansal2017,An2018,ChengSohn2024}), and the only one below $1~{\rm pc}$, and places the binary of NGC~5044 in the regime of bound binaries.

\section{Discussion}\label{sec:discussion}
\subsection{History of AGN activity in NGC~5044}\label{subsec:historyAGN}
\begin{figure}[htp!]
    \centering
    \includegraphics[width=\linewidth]{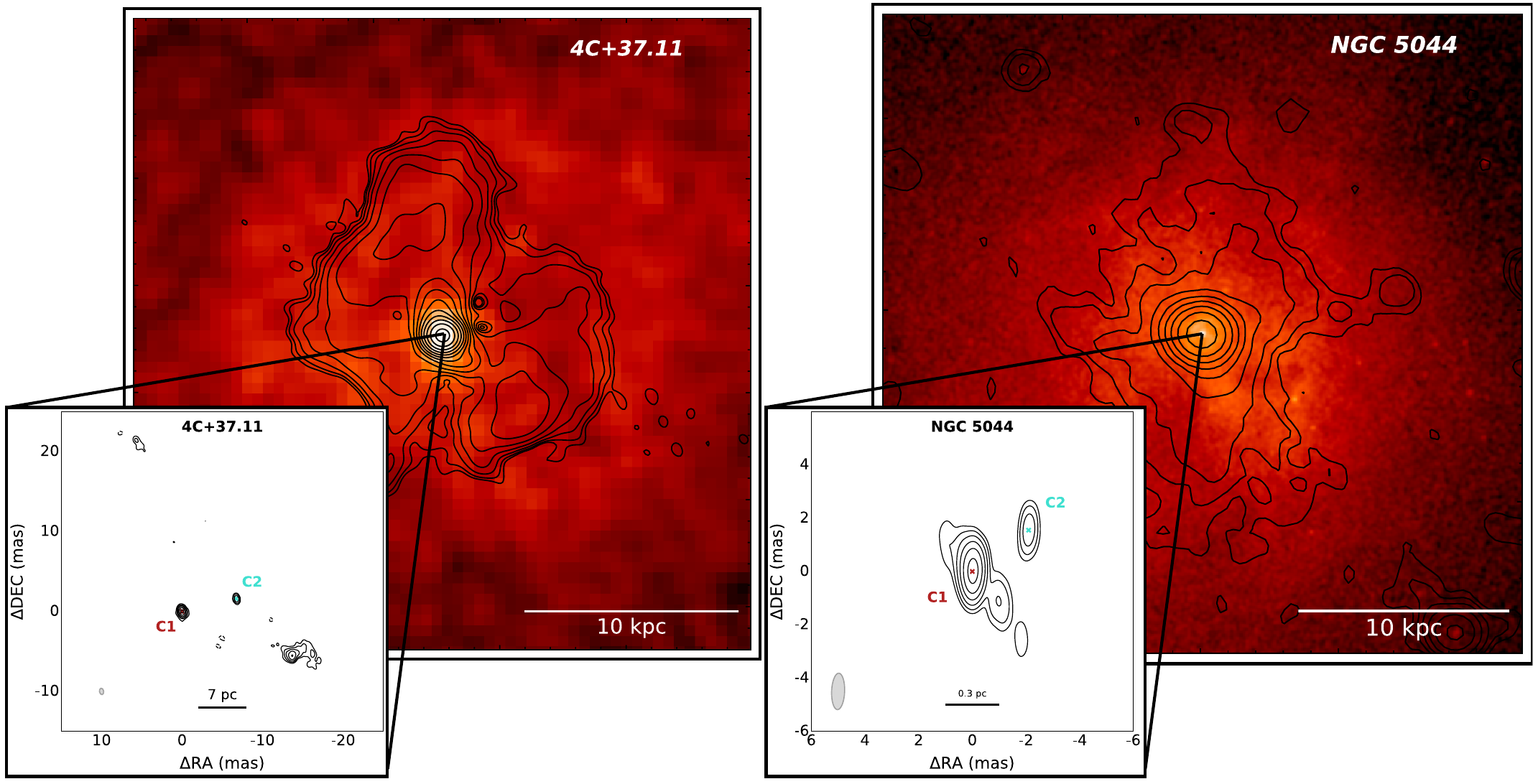}
    \caption{Comparison between the two known pc-scale binary SMBH systems in central cluster ellipticals. Main panels: \textit{Chandra} 0.5--2~keV images of 4C+37.11 (left) and NGC~5044 (right), showing the surrounding hot intracluster/intragroup gas, overlaid with radio contours from the VLA at 1.5 GHz (4C+37.11, \cite{Romani2014}) and uGMRT at 400 MHz (NGC~5044, \cite{Rajpurohit2025}). Insets: VLBA 15~GHz images resolving each binary, from \cite{Bansal2017} (4C+37.11) and this work (NGC~5044). Observational parameters are given in {\bf Methods}.
    }
    \label{fig:comparison}
\end{figure}

The discovery of a sub-pc binary of active SMBHs in NGC~5044 has direct implications for the puzzle of the activity history of its central AGN. Deep \textit{Chandra} X-ray data revealed two pairs of cavities, or bubbles, excavated in the hot intragroup gas by past radio outbursts, both aligned along a northwest-southeast axis, with the older, larger pair estimated to be $\approx$80~Myr old \cite{Schellenberger2021,Rajpurohit2025} (Figure~\ref{fig:comparison}, right panel), and the inner, smaller pair being $\approx$1~Myr old \cite{David2017,Schellenberger2021}. However, the jets currently launched by the central AGN, as imaged with the VLBA (Figure~\ref{fig:comparison}, inset), are misaligned by $\sim$90$^{\circ}$ with respect to the kpc-scale cavities \cite{Schellenberger2021,Ubertosi2026}. Such misalignments between pc- and kpc-scale radio structures are not uncommon in cluster- and group-central galaxies, occurring in about 30\% of cases, although their physical origin remains poorly understood \cite{Ubertosi2024}. NGC~5044 stands out as the most extreme example identified to date: the relative ages of the young VLBA-scale outburst and the inner X-ray cavities imply that the jet must have reoriented in less than 1~Myr \cite{Schellenberger2021,Ubertosi2024} -- a timescale difficult to reconcile with single-SMBH precession and reorientation models (see also \cite{Ubertosi2021}), but compatible with a scenario where the galaxy hosts two SMBHs independently launching jets.

The binary system reported here offers a solution to this mismatch between pc-scale and kpc-scale jets: without requiring any jet reorientation, the kpc-scale cavities may instead have been excavated by a past outburst from the secondary core (C2), leaving the jets currently launched by the primary core (C1) unchanged in orientation. This scenario may be supported by the detection of a “radio spur" extending northwest of the core \cite{Ubertosi2026}, also visible in the 4.9~GHz image in Figure~\ref{fig:imagesVLBA} (top left), which aligns with the larger-scale cavities in the intragroup medium. It is possible that this spur is the relic of the jets that excavated the multiple generations of kpc-scale X-ray cavities.

\subsection{Central elliptical galaxies and binary SMBH systems}\label{subsec:clusterbinaryconnection}
The only secure, directly imaged, pc-scale confirmed systems of binary SMBHs is hosted in the galaxy 4C+37.11. Interestingly, this system also inhabits the central elliptical galaxy of a galaxy cluster. This makes it particularly interesting to compare 4C+37.11 with NGC~5044, as both systems share a number of similarities (see Figure~\ref{fig:comparison}). 

First, the merger histories of both galaxies suggest a long period of dynamical relaxation. According to \cite{Mendel2008}, the group of NGC~5044 has experienced no major subgroup merger for at least the past 1~Gyr. This is consistent with the stellar population analysis of \cite{Diniz2017}, who find dominant stellar ages of 900~Myr (40\% of the stars) and 13~Gyr for the central galaxy. Similarly, for 4C+37.11, \cite{andradesantos2016} report no evidence for a major merger within the last $\geq 2$~Gyr. Such timescales are sufficient for supermassive black holes (SMBHs) originating from earlier mergers to in-spiral and reach the galactic center.

Second, both galaxies exhibit more massive black holes than expected from their stellar velocity dispersions. Galaxies that primarily grow via dry mergers (as cluster or group central galaxies, e.g., \cite{Liu2009,Lavoie2016}) tend to develop cored stellar profiles (e.g., \cite{Mehrgan2019,McDonald2026}); in such systems the stellar velocity dispersion increases more slowly than the black hole mass. For NGC~5044, dynamical modeling of the ionized gas in the innermost regions \cite{David2017} and modeling of the AGN spectral energy distribution \cite{Schellenberger2021} yield a (total) black hole mass of $\sim2\times10^{9}$~M$_{\odot}$, an order of magnitude larger than inferred from the $M_{\mathrm{BH}}-\sigma$ relation \cite{David2009}. The case is similar for 4C+37.11, where the total SMBH mass (C1 + C2) is $M_{\mathrm{tot}}^{\mathrm{BH}} = 2.8\times10^{10}$~M$_{\odot}$, about seven times higher than predicted by its stellar velocity dispersion \cite{Surti2024}. 

These pieces of evidence, together with the location of both systems at the centers of dynamically relaxed group or cluster environments (see \cite{Romani2014,andradesantos2016} for 4C+37.11), are consistent with the expectation that massive, radio-loud central elliptical galaxies are the most favorable sites for detecting SMBH binaries with VLBI. This picture is also possibly supported by the case of GPS~J1543-0757, another recently identified candidate binary SMBH system, with a larger separation of 130~pc, also imaged in radio with VLBI \cite{ChengSohn2024} and also hosted in an elliptical galaxy \cite{Mahony2011}.

\clearpage

\section{Methods}\label{sec:methods}

\subsection{Radio data reduction and analysis: VLBI observations}\label{subsec:vlbi-reduction}
We analyze archival and proprietary VLBA observations of NGC~5044 (see Table \ref{tab:data} for full details) from March 2020 (11h, project BS~283 at 4.9 \& 8.4 GHz, PI Schellenberger), May-June 2022 (6h, project BS~301 at 23.6 GHz, PI Schellenberger), September 2024 (14h, project BU~039 at 4.9 \& 8.4 GHz, PI Ubertosi), and May 2026 (8h, project BU~042 at 15.2 GHz, PI Ubertosi). The observations were performed in phase referencing mode, using the calibrator source J1317-1345 
during projects BS~283 and BU~039, and using the calibrator source J1306-1718 
during projects BS~301 and BU~042. The observations were reduced using standard techniques in AIPS \cite{Greisen2003}, with a correction of Earth orientation parameters (VLBAEOPS) and ionospheric delays (VLBATECR). Then, we solved for phase delays and amplitudes by applying digital sampling corrections (VLBACCOR), by removing the effect of instrumental delays on phases (VLBAMPCL), and by calibrating the bandpass (BPASS). We fringe-fitted the data (FRING) and transferred the solution from the phase calibrator to the target (CLCAL). The phases of the phase calibrator were further refined with self-calibration (tasks CALIB and IMAGR) before transferring again the solution to the target. Then, we applied all the calibration to the target data, averaged channels within the different spectral windows, and split out a corrected dataset (SPLIT). 
\par Subsequently, we performed phase and amplitude self-calibration of the target. For the observations at 4.9 GHz (C band) and 8.4 GHz (X band) from BS~283 and BU~039, five cycles of self-calibration were performed (four phase-only + one amplitude and phase), until there was no improvement in rms noise or dynamic range of the final images. For the observations at 23.6 GHz (K band) from BS~301, four cycles of phase-only self calibration were performed after averaging together the different spectral windows. The rms noise of our final images at 23.6 GHz is a factor of 2 worse than expected, likely because the phase calibrator showed weak fringes during the BS~301 observations. Although fringes were detected at 23.6 GHz and calibration was possible, data were also taken at 43 GHz (Q band) during project BS~301, but could not be calibrated due to this fringe-detection issue. For the 15.2 GHz (U band) observations from BU~042, after two cycles of phase-only self-calibration the image quality and rms noise already reached the levels expected from the data. 

\begin{table}[ht!]
     \centering
     \caption{Summary of VLBA observations and images employed in this work. }\label{tab:data}\renewcommand*{\arraystretch}{1.2}
     \begin{tabular}{ccccc}
     \hline
 Obs. band & Obs. Date & t$_{obs}$  & Calibrators  & UV range  \\
 (1) & (2) & (3) & (4) & (5) \\
\hline
 \multicolumn{5}{c}{{\bf Project BS283}} \\

  \multirow{1}{*}{C  (4.9 GHz)} & \multirow{1}{*}{Mar. 2020} & \multirow{1}{*}{5.5h} & \multirow{1}{*}{J0927+3902 (FF), J1317-1345 (PC)}  & 2 -- 150 M$\lambda$\\

  \multirow{1}{*}{X  (8.4 GHz)} & \multirow{1}{*}{Mar. 2020} & \multirow{1}{*}{5.5h} & \multirow{1}{*}{J0927+3902 (FF), J1317-1345 (PC)} & 5 -- 230 M$\lambda$\\

 \multicolumn{5}{c}{{\bf Project BS301}} \\

 \multirow{1}{*}{K  (23.6 GHz)} & \multirow{1}{*}{May 2022} & \multirow{1}{*}{6h} & \multirow{1}{*}{J0927+3902 (FF), J1306-1718 (PC)} & 20 -- 700 M$\lambda$ \\

 \multicolumn{5}{c}{{\bf Project BU039}} \\

  \multirow{1}{*}{C  (4.9 GHz)} & \multirow{1}{*}{Sep. 2024} & \multirow{1}{*}{7h} & \multirow{1}{*}{3C286 (FF), J1317-1345 (PC)} & 2 -- 140 M$\lambda$ \\

 
  \multirow{1}{*}{X  (8.4 GHz)} & \multirow{1}{*}{Sep. 2024} & \multirow{1}{*}{7h} & \multirow{1}{*}{3C286 (FF), J1317-1345 (PC)} & 6 -- 300 M$\lambda$ \\

\multicolumn{5}{c}{{\bf Project BU042}} \\

 \multirow{1}{*}{U  (15.2 GHz)} & \multirow{1}{*}{May 2026} & \multirow{1}{*}{8h} & \multirow{1}{*}{3C345 (FF), J1306-1718 (PC)} & 8 -- 450 M$\lambda$ \\

  \hline

\end{tabular}
\footnotetext{(1) Observing band and central frequency; (2) date of the observations; (3) total time; (4) Calibrators (FF = fringe finder, PC = phase calibrator); (5) UV range.}
 \end{table}

 \begin{table}[htp!]
\centering
\caption{Imaging parameters for the VLBA maps shown in Figure~\ref{fig:imagesVLBA}. }\label{tab:imgparams}
\begin{tabular}{cccccc}
\hline
Frequency & Epoch & Robust & $\sigma_{\rm rms}$ (mJy/beam) & Beam FWHM & PA \\
(1) & (2) & (3) & (4) & (5) & (6) \\
\hline
4.9 GHz  & 2024 & +2 & 0.010 & $4.89\times1.90$ mas & $+13.0^\circ$ \\
8.4 GHz  & 2024 & $-1$ & 0.028 & $1.96\times0.75$ mas & $-7.6^\circ$ \\
15.2 GHz & 2026 & 0 & 0.029 & $1.37\times0.49$ mas & $-2.13^\circ$ \\
23.6 GHz & 2022 & 0 & 0.057 & $1.26\times0.38$ mas & $-13.6^\circ$ \\
\hline
\end{tabular}
\footnotetext{Columns: (1) Central frequency; (2) epoch of the observations (3) Robust parameter adopted for Briggs weighting; (4) rms noise of the image; (5) restoring beam FWHM; (6) restoring beam PA.}
\end{table}

    \begin{table}[t]
     \centering
     \caption{Flux density and position of the \textit{(u,v)}-plane components identified in the data listed in Table~\ref{tab:data}. An inner uv-cut of $30$~M$\lambda$ was imposed to the datasets to ensure emission is recovered on the same angular scales.}\label{tab:difmap}
     \begin{tabular}{ccccccccc}

     \hline     
     & $\nu$ & Epoch & $S_{\nu}$ & SNR & $\Delta$RA & $\Delta$DEC & FWHM & $r_{proj,c}$  \\
    &  GHz & & mJy & & mas & mas & mas & mas  \\
   (1) & (2) & (3) & (4) & (5) & (6) & (7) & (8) & (9) \\
 
\hline

     \multirow{6}{*}{C1} & \multirow{2}{*}{4.9}  & 2020& $5.90\pm0.60$ & 98 & 0 & 0&   0.36 & 0 \\

    & & 2024& $3.29\pm0.33$ & 65 & 0 & 0& 0.11  &  0 \\
    
  & \multirow{2}{*}{8.4} & 2020 & $7.90\pm0.80$ &  136 & 0 & 0 & 0.44 & 0  \\
  & & 2024 & $5.61\pm0.56$ & 125 & 0 & 0 & 0.41 &  0 \\
  & 15.2 & 2026 & $6.53\pm0.65$ & 177  & 0 & 0 & 0.22  &  0  \\
& 23.6 & 2022 & $7.32\pm0.75$ & 90 & 0 & 0 & 0.16 &  0 \\
\hline

 \multirow{6}{*}{C2} & \multirow{2}{*}{4.9}  & 2020& $0.80\pm0.10$ & 13 & $-2.05$ & $+2.07$ & 0.33 &  $2.92\pm0.15$  \\

    & & 2024& $0.70\pm0.09$ & 14 & $-2.43$ & $+0.88$ & 0.22 &  $2.58\pm0.15$   \\
    
  & \multirow{2}{*}{8.4} & 2020 & $1.32\pm0.14$ & 23  & $-2.18$ & $+1.53$&  0.30 &  $2.66\pm0.07$  \\
  & & 2024 & $0.98\pm0.11$ & 22 & $-2.10$ & $+1.54$& 0.42 & $2.61\pm0.06$ \\

  & 15.2 & 2026 & $1.31\pm0.14$ &  35 & $-2.06$ & $+1.52$& 0.18 &  $2.56\pm0.02$  \\
& 23.6 & 2022 & $1.12\pm0.13$ & 13  & $-2.03$ & $+1.66$& 0.26 & $2.62\pm0.04$ \\


\hline
\multirow{6}{*}{SW jet} & \multirow{2}{*}{4.9}  & 2020& $6.02\pm0.60$ & 100 & $-1.36$ & $-1.46$& 1.67 & $1.99\pm0.02$\\

    & & 2024& $6.30\pm0.63$ & 124 & $-0.87$ & $-1.38$& 2.01 &  $1.63\pm0.02$ \\
    
  & \multirow{2}{*}{8.4} & 2020 & $2.78\pm0.28$ & 46  & $-1.18$ & $-0.90$& 0.84 & $1.49\pm0.03$ \\
  
  & & 2024 & $3.31\pm0.33$ & 73 & $-0.49$ & $-1.14$& 0.58  &  $1.25\pm0.02$  \\

  & 15.2 & 2026 & $1.26\pm0.13$ & 34  & $-0.94$ & $-1.07$& 0.70 &  $1.48\pm0.02$  \\
& 23.6 & 2022 & $\leq0.53$ & -- & -- & -- & -- &  -- \\
  \botrule
\end{tabular}
\footnotetext{Columns: (1) Component; (2) observing frequency; (3) epoch; (4) flux density and associated uncertainty, calculated as $\delta S_{\nu} = \sqrt{(10\%S_{\nu})^2 + N_{\rm beam}\sigma_{rms}^{2}}$, where $N_{\rm beam}$ is the ratio between the area within which the flux
density is computed and the beam area; (5) signal-to-noise ratio of the component; (6) offset in right ascension from the core C1; (7) offset in declination from the core C1; (8) extent (FWHM) of the circular Gaussian used to fit the component; (9) projected distance of the component from the core C1 and associated uncertainty, corresponding to the beam FHWM (see Figure~\ref{fig:uvcutimages}) divided by the SNR of the components.}
 \end{table}

\paragraph{Imaging and identification of components}\label{par:imaging}
Using the final calibrated datasets at 4.9 GHz (2024), 8.4 GHz (2024), 15.2 GHz (2026), and 23.6 GHz (2022), we produced the images shown in Figure~\ref{fig:imagesVLBA} using the software CASA \cite{CASATeam2022} and adopting Briggs weighting \cite{Briggs1995} and multi-scale cleaning (with scales of 0, 1, and 4 synthesized beams). To highlight structures at different scales and with different angular resolutions, we adopted Briggs robustness parameters of \texttt{R=+2} at 4.9 GHz, \texttt{R=-1} at 4.9 GHz, \texttt{R=0} at 15.2 GHz, and \texttt{R=0} at 23.6 GHz. Other choices of Briggs robustness (i.e., more natural or more uniform imaging) at 15.2 GHz and 23.6 GHz do not affect the detection of C1 and C2; the choice of \texttt{R=0} simply offers the best compromise between angular resolution and sensitivity. The rms noise levels and angular resolution (beam FWHM) are reported in Table~\ref{tab:imgparams}.
The images in Figure~\ref{fig:imagesVLBA} highlight the northeast-southwest symmetric jet structure extending from the core C1 (which was already discussed and analyzed in \cite{Schellenberger2021,Ubertosi2024,Ubertosi2026}), and further spatially resolve (at 8.4 GHz, 15.2 GHz, and 23.6 GHz) and detect at high SNR (see Table~\ref{tab:difmap}) the secondary compact core C2.
For the analysis presented below, the source structure was characterized directly in the visibility domain.

\par
\begin{figure}[htp!]
    \centering
    \includegraphics[width=0.32\linewidth]{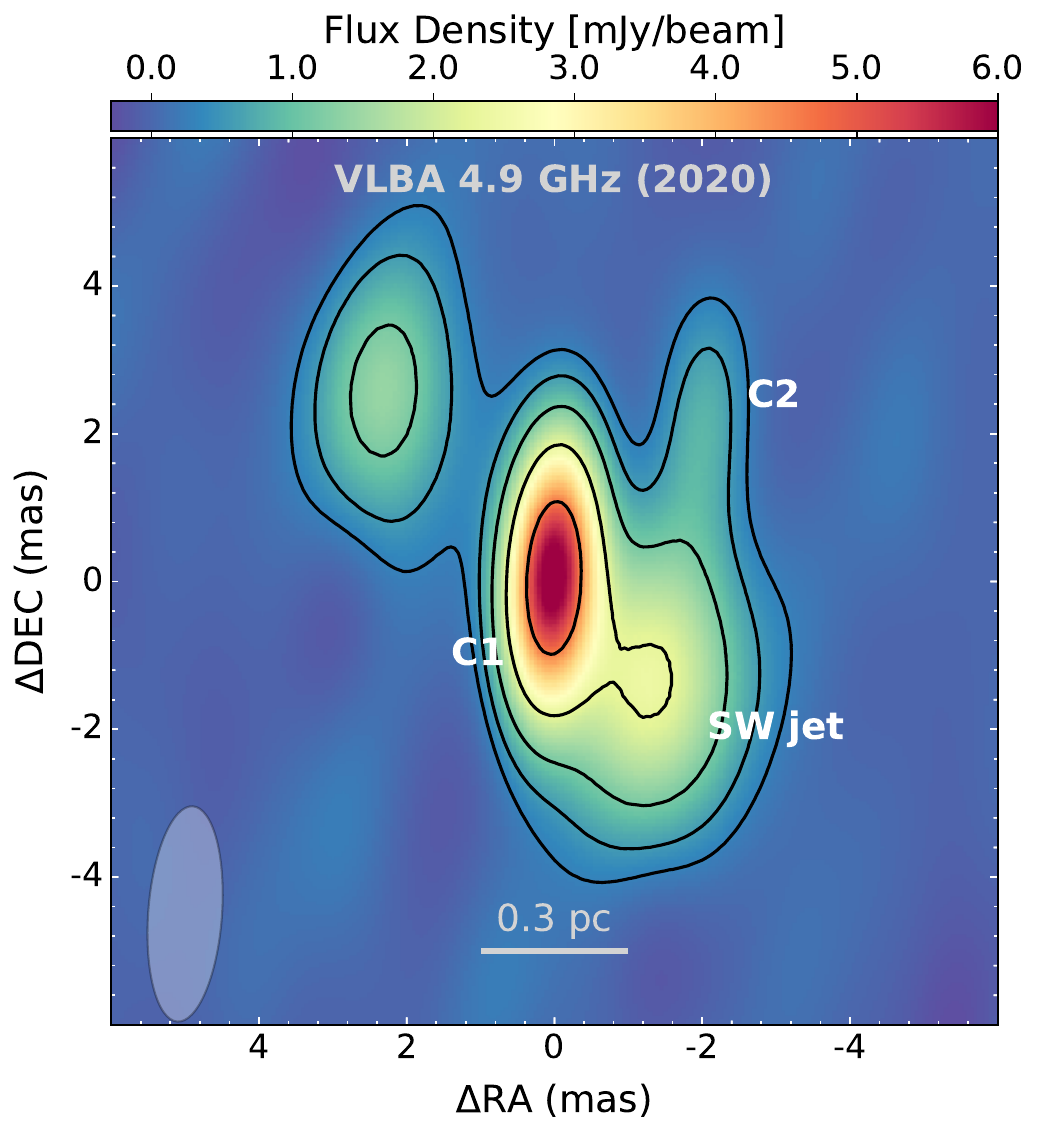}
    \includegraphics[width=0.32\linewidth]{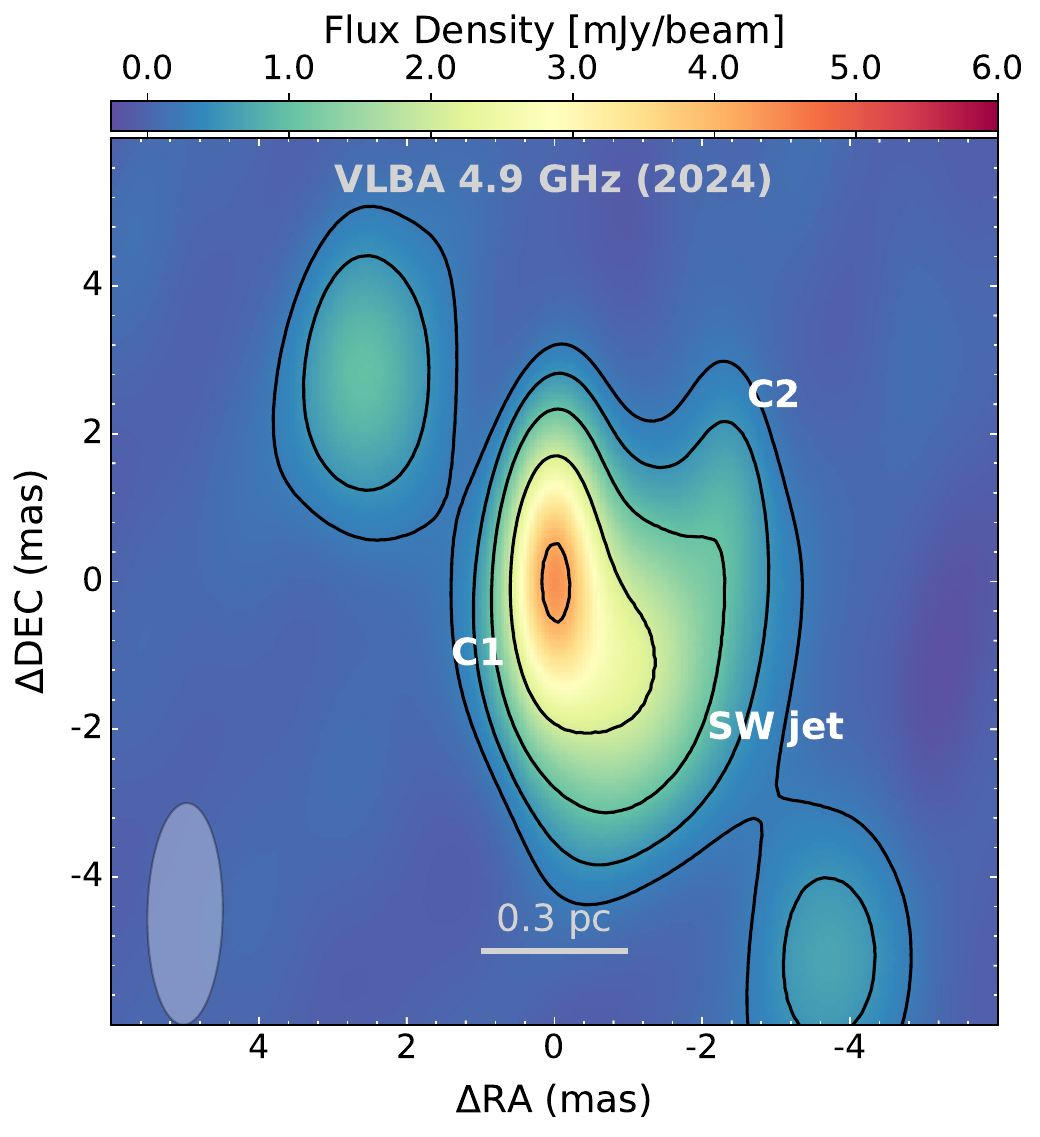}
    \includegraphics[width=0.32\linewidth]{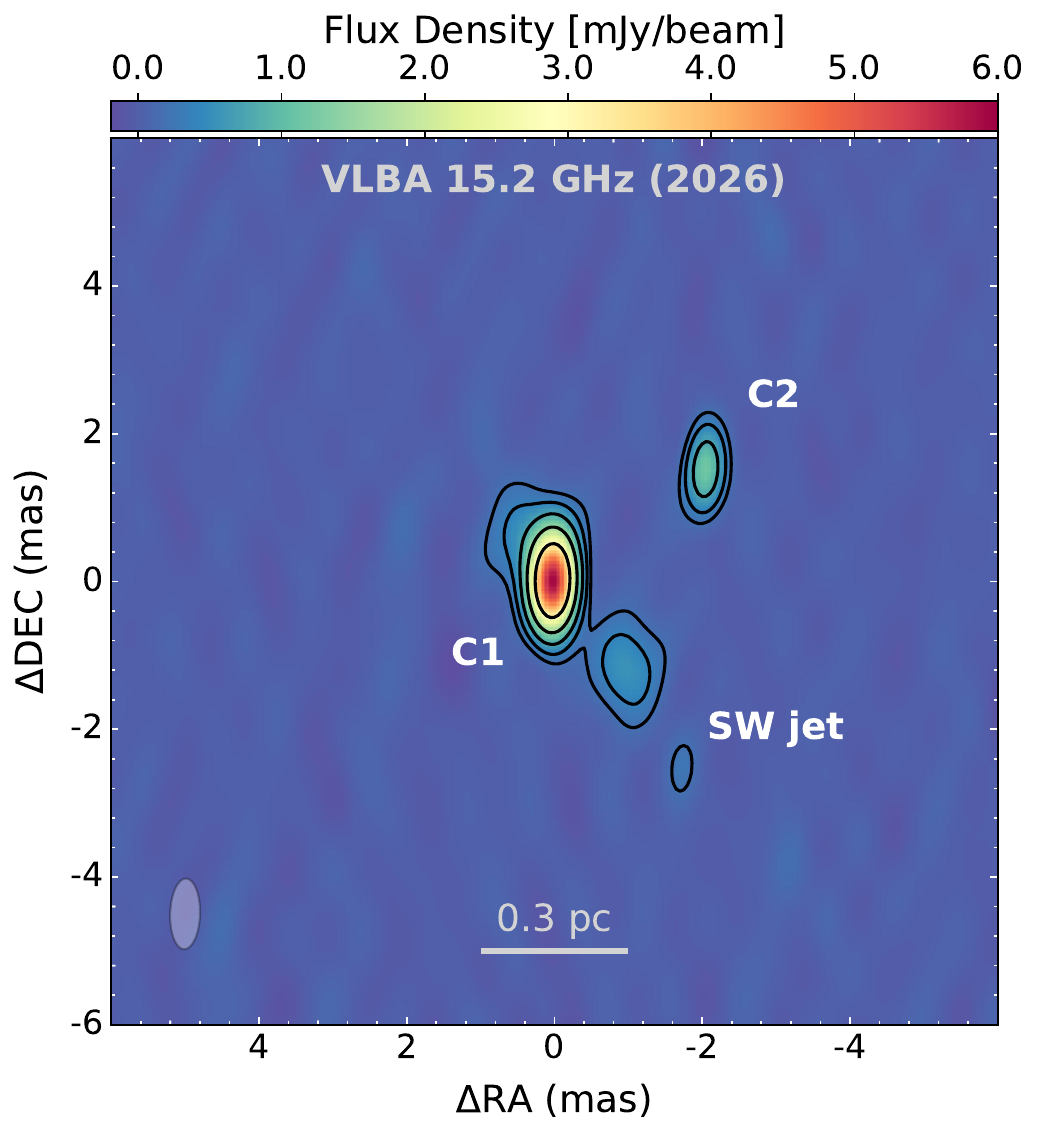}\\
    \includegraphics[width=0.32\linewidth]{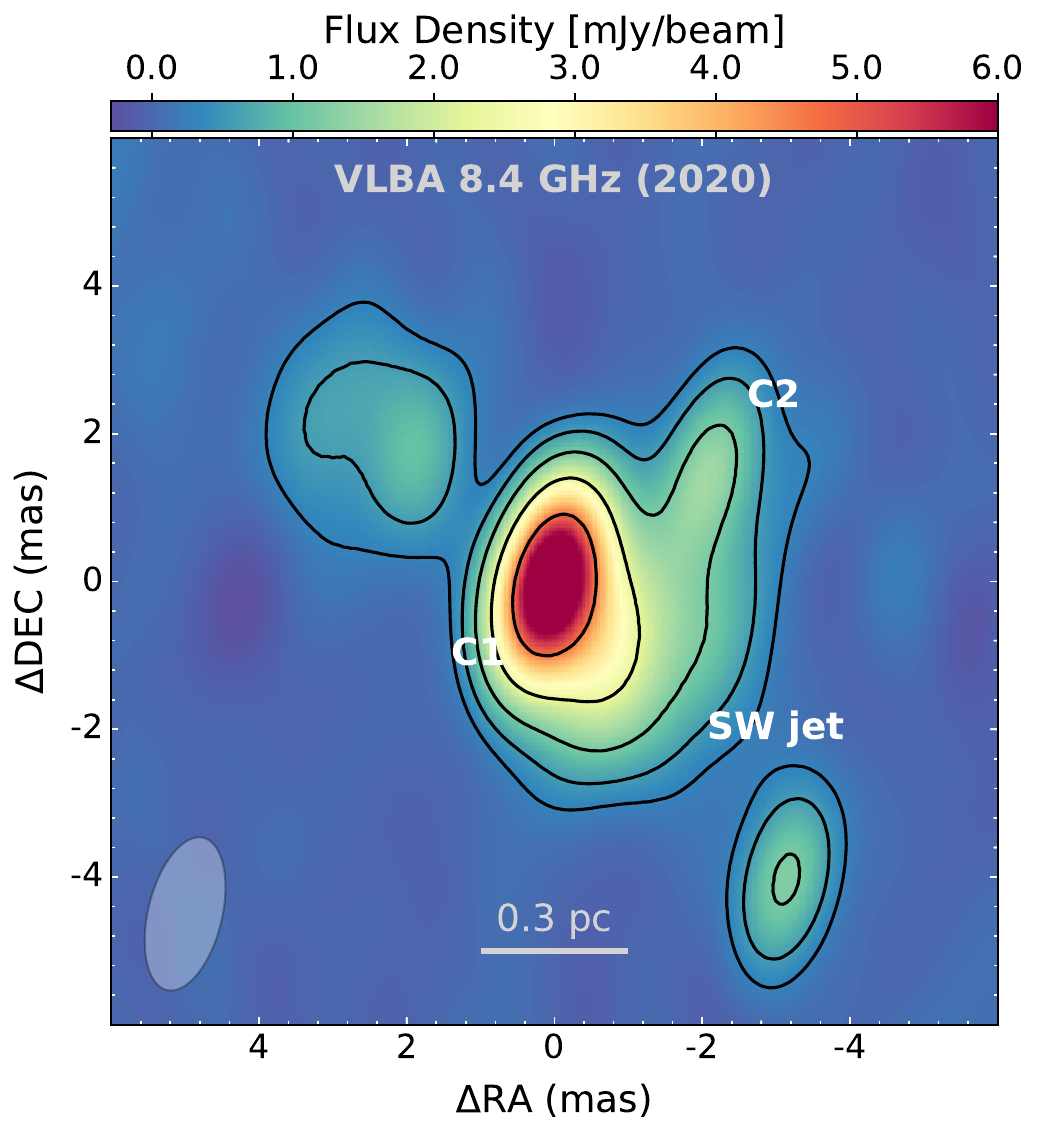}
    \includegraphics[width=0.32\linewidth]{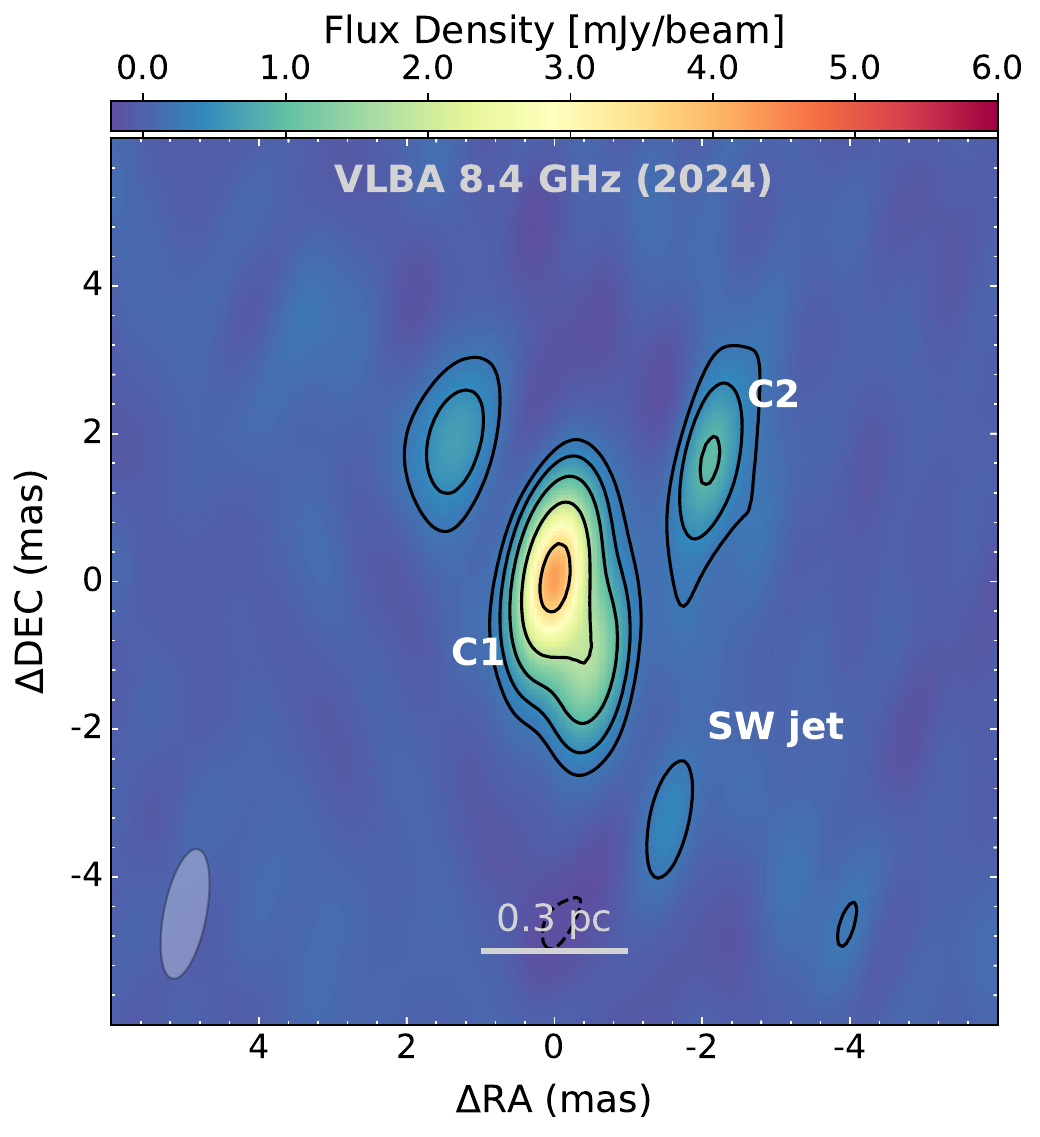}
    \includegraphics[width=0.32\linewidth]{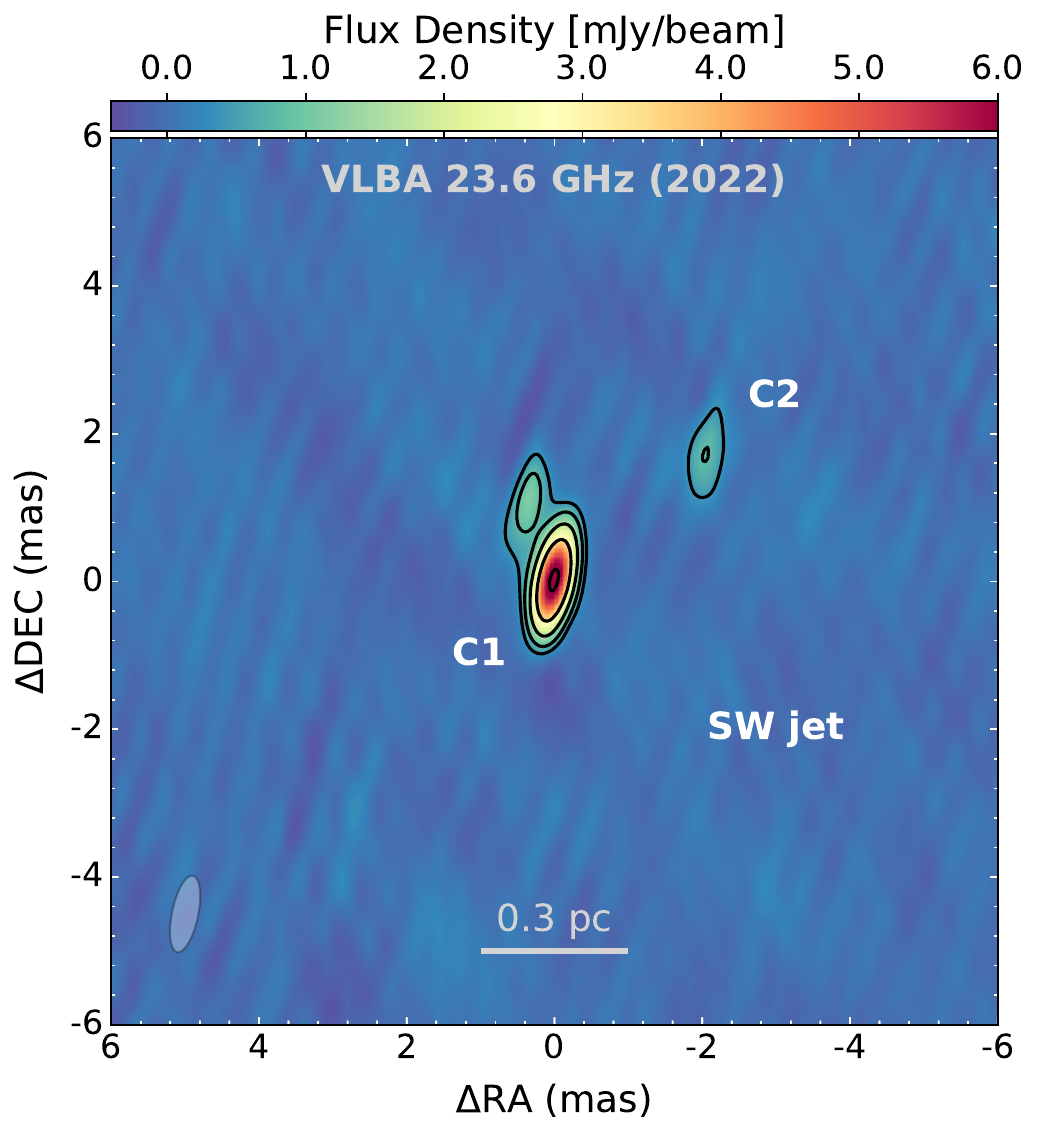}\\
    \caption{
    Multi-wavelength VLBA images of the core of the galaxy NGC~5044, obtained in \texttt{Difmap} after imposing an inner uv-cut of 30~M$\lambda$ to the visibilities to obtain uniform sensitivity to extended structures. Images are labelled with the corresponding frequency and date. In all panels, black contours are drawn starting at 5$\times\sigma_{rms}$ (with rms noises of $60\,\mu{\rm Jy\,beam^{-1}}$ at 4.9 GHz, 2020; $51\,\mu{\rm Jy\,beam^{-1}}$ at 4.9 GHz, 2024; $58\,\mu{\rm Jy\,beam^{-1}}$ at 8.4 GHz, 2020; $45\,\mu{\rm Jy\,beam^{-1}}$ at 8.4 GHz, 2024; $38\,\mu{\rm Jy\,beam^{-1}}$ at 15.2 GHz, 2026; $81\,\mu{\rm Jy\,beam^{-1}}$ at 23.6 GHz, 2022;) and increase by factors of two. The first negative contour at $-5\times\sigma_{rms}$ is dashed. The angular resolution of the image (beam FWHM) is shown with an ellipse in the bottom left corner ($2.92\,{\rm mas}\times1.00\,{\rm mas}\,,-4.14^{\circ}$ at 4.9~GHz, from 2020; $3.00\,{\rm mas}\times1.02\,{\rm mas}\,,-0.99^{\circ}$ at 4.9~GHz, from 2020; $2.12\,{\rm mas}\times1.00\,{\rm mas}\,,-13.5^{\circ}$ at 8.4~GHz, from 2020; $1.79\,{\rm mas}\times0.58\,{\rm mas}\,,-11.0^{\circ}$ at 8.4~GHz, from 2024; $0.96\,{\rm mas}\times0.40\,{\rm mas}\,,-1.68^{\circ}$ at 15.2~GHz, from 2026; $1.05\,{\rm mas}\times0.35\,{\rm mas}\,,-11.8^{\circ}$ at 23.6~GHz, from 2022). The two compact cores C1 and C2, as well as the southwestern jet knot for which we report the flux densities in Table~\ref{tab:difmap} (and plotted in Figure~\ref{fig:spectracomponents}), are labeled in white.
    }
    \label{fig:uvcutimages}
\end{figure}

 We used the software package \texttt{Difmap} \citep{Shepherd1997} to model the calibrated visibilities with circular Gaussian components. This approach avoids biases introduced by image weighting and deconvolution, and is the standard approach in VLBI studies of compact radio sources (e.g., \cite{Giroletti2003,Giroletti2005,Rodriguez2006,Bansal2017}). To ensure comparable sensitivity to compact and extended emission across the different datasets, we imposed a common lower baseline cut of 30~M$\lambda$. This threshold was chosen by comparing the $(u,v)$ coverage of the various observations (see last column in Table~\ref{tab:data}), and filters out structures larger than approximately 7~mas. We modeled the emission associated with the primary core (C1), the secondary compact component (C2), and the south-west jet, all of which are detected in most datasets. The resulting model parameters are reported in Table~\ref{tab:difmap} (see also Figure~\ref{fig:uvcutimages}).

\par No compact counterpart of the SW jet is detected at 23.6~GHz. 
We estimated a $5\sigma$ upper limit for this component by combining the rms noise in the residual K-band image (0.08~mJy~beam$^{-1}$) with the expected area of the jet component, which was in turn estimated from the Gaussian model at 15.2~GHz (FWHM = 0.7~mas) -- the closest frequency at which the component is detected -- and converted into an equivalent number of restoring beams at 23.6~GHz (1.05~mas $\times$ 0.35~mas). This procedure yields a $5\sigma$ upper limit of 0.53~mJy (Table~\ref{tab:difmap}). Comparing this upper limit with measurements at lower frequencies (Figure~\ref{fig:spectracomponents}, right) suggests the presence of a spectral break between 10~GHz and 20~GHz, as expected for synchrotron-emitting plasma undergoing radiative losses.

\paragraph{Frequency-dependent core position}\label{par:core-shift}
The apparent separation between C1 and the SW jet decreases systematically with increasing observing frequency. This behaviour is expected from the synchrotron core-shift effect \citep{Lobanov1998,Pushkarev2012}: the observed VLBI core corresponds to the surface with an opacity $\tau_\nu \simeq 1$, and therefore moves towards the central engine at higher frequencies. We note that in NGC~5044 this frequency dependence is superimposed on the proper motion of individual jet knots (studied in \cite{Ubertosi2026}), making the SW jet unsuitable for a quantitative determination of the intrinsic core shift. Coeval multifrequency observations would be needed to study this effect.

In contrast, the relative position of C2 with respect to C1 is expected to be less affected. The jets emerging from C1 are oriented approximately perpendicular to the C1--C2 axis (Figure~\ref{fig:imagesVLBA}), so any frequency-dependent displacement of the C1 core occurs largely orthogonally to the measured binary separation. Indeed, the projected C1--C2 distance remains nearly constant over the full frequency range, showing a mild increase towards lower frequencies. Fitting the measured separations with the standard parametrization (e.g., \cite{Lobanov1998}):
\begin{equation}
    r_{\rm proj}(\nu)=A+B\nu^{-1},
\end{equation}
and assuming the FWHM of the restoring beam divided by the SNR as the uncertainties on the distance $r_{\rm proj}$ (e.g., see \cite{Ubertosi2026}; see values in Table~\ref{tab:difmap}) we obtain:
\begin{equation}
    r_{\rm proj}(\nu)= (2.52\pm0.10)+(0.88\pm0.55)(\nu{\rm\,[GHz]})^{-1}\ {\rm mas},
\end{equation}
with $\chi^{2}/{\rm d.o.f.} = 1.59$. Extrapolating to infinite frequency gives an intrinsic projected separation of:
\begin{equation}
    r_{\rm proj}(\infty)=2.52\pm0.10~{\rm mas},
\end{equation}
corresponding to $0.38\pm0.02$~pc. We thus adopt this value as the best estimate of the projected separation between the two compact cores.

\paragraph{Radio spectra}\label{par:spectra}
Figure~\ref{fig:spectracomponents} shows the radio spectra of C1, C2, and the SW jet. Both compact components exhibit flat to inverted spectra ($\alpha \gtrsim 0$), consistent with partially self-absorbed AGN cores (e.g., \cite{Rodriguez2006,Hovatta2014}). In addition, the flux densities of both C1 and C2 vary between the 2020 and 2024 observations at 4.9 and 8.4~GHz, suggesting intrinsic variability -- another typical property of radio AGN cores (e.g., \cite{Lister2009,Hogan2015}). We note that in the spectral index maps at lower angular resolution ($\approx3$~mas) presented in \cite{Schellenberger2021,Ubertosi2026}, the region of flatter spectral index was not in the center of the core, but offset to northeast and southwest of the core. This is most likely caused by the presence of the two flat spectrum compact cores C1 and C2, blended into one component at lower angular resolution.

The SW jet instead displays the steep spectrum characteristic of optically thin synchrotron emission. We also note that its flux density at 4.9 GHz and 8.4 GHz remained very similar between 2020 and 2024. By measuring the spectral index between 4.9 GHz and 8.4 GHz at both epochs, we find $\alpha = -1.2\pm0.3$, consistent with an optically thin AGN jet (e.g., \cite{Hovatta2014}).

\subsection{Comparison between 4C+37.11 and NGC~5044}
To place the pc-scale binary in NGC~5044 in a broader context, we compare it with 4C+37.11, the only confirmed pc-scale binary AGN directly imaged with VLBI  \cite{Rodriguez2006,Bansal2017} -- also hosted in a central elliptical galaxy. For this comparison, we re-reduce and use archival radio observations of 4C+37.11 and archival X-ray observations of both systems. For NGC~5044, we also overlay on Figure~\ref{fig:comparison} the uGMRT Band 3 contours an angular resolution of $10''$ presented by \cite{Rajpurohit2025}, starting at $3\times\sigma_{rms}$ (with rms noise of $21\,\mu{\rm Jy\,beam^{-1}}$) and increasing by a factor of 2.

\paragraph{Radio data: 4C+37.11}
We retrieved archival 15~GHz VLBA observations of 4C+37.11 from 2015 (project BT~129), originally presented in \cite{Bansal2017}. These data were calibrated following the same procedure adopted for the U-band (15.2~GHz) dataset of NGC~5044 described above. The final calibrated visibilities were imaged in CASA using Briggs weighting (\texttt{R=0}) and multi-scale clean (scales of 0, 1, and 4 beams), yielding the image shown in the left subpanel of Figure~\ref{fig:comparison}, with an angular resolution of $0.83\,{\rm mas}\times0.54\,{\rm mas}, +10.0^{\circ}$ and rms noise of $220\,\mu{\rm Jy\,beam^{-1}}$.

We also retrieved 1.5 GHz VLA observations of 4C+37.11 from 2012 (project 12B-058), originally presented in \cite{Romani2014}. We obtained the fully calibrated visibilities from the NRAO Archive, and subsequently performed self-calibration on the target in CASA. The final calibrated visibilies were imaged in CASA using Briggs weighting (\texttt{R=0}), obtaining a final image with an angular resolution of $0.92''\times0.82''$. Contours were generated from this image and overlaid on Figure~\ref{fig:comparison} (left panel), starting at $3\times\sigma_{rms}$ (with rms noise of $30\,\mu{\rm Jy\,beam^{-1}}$) and increasing by a factor of 2.

\paragraph{\textit{X-ray} data: 4C+37.11 and NGC~5044}
We reprocessed archival \textit{Chandra} ACIS observations of 4C+37.11 (ObsID 16120) and NGC~5044 (ObsIDs 798, 9399, 17195, 17196, 17653, 17654) following the same procedure described in \cite{Ubertosi2024}. In particular, the event files were reprocessed with the software CIAO \cite{Fruscione2006} to apply the latest calibration products. Time intervals affected by anomalously high count rates were filtered out. Astrometric registration of each observation was refined by cross-matching detected X-ray point sources against optical catalogs, and, for NGC~5044, individual event files were merged. We then produced the exposure-corrected, background-subtracted images in the 0.5--2~keV band shown in the main panels of Figure~\ref{fig:comparison} (where we applied a light Gaussian smoothing on scales of 1.5~arcsec). These images highlight the diffuse hot intracluster/intragroup gas surrounding each galaxy.

\bmhead{Acknowledgements}
FU acknowledges the financial contribution from contract PRIN 2022 - CUP J53D23001610006. Basic research in radio astronomy at the Naval Research Laboratory is supported by 6.1 Base funding. The National Radio Astronomy Observatory and Green Bank Observatory are facilities of the U.S. National Science Foundation operated under cooperative agreement by Associated Universities, Inc.

\bibliography{sn-bibliography}

\end{document}